\documentclass[final,5p,times,twocolumn]{elsarticle}
\usepackage{graphicx}
\usepackage{amssymb}
\usepackage{amsmath}

\newcommand{\myvec}[1]{#1}

\begin{document}

\begin{frontmatter}

\title{Encryption dynamics and avalanche parameter for ``delayed dynamics''-based cryptosystems}

\author{O. Melchert}
\ead{oliver.melchert@uni-oldenburg.de}

\address{
Institut f\"ur Physik, Universit\"at Oldenburg, 
Carl-von-Ossietzky Stra\ss{}e 9--11, 26111 Oldenburg, Germany} 

\begin{abstract}
The presented article attempts to characterize the encryption dynamics of delayed dynamics based
block ciphers, designed for the encryption of binary data.
For such encryption algorithms, the encryption process relies on a coupling dynamics 
with time delay between different bits in the plaintext (i.e.\ the 
``initial'' message to be encrypted).
Here, the principal dynamics of the encryption process is examined and the 
Hammingdistance is used to quantify the change in ciphertext (i.e.\ the plaintext after encryption) 
upon changing a single bit in the plaintext message or slightly perturbing the key used during 
encryption.
More precisely, the previously proposed ``encryption via delayed dynamics'' (in short: EDDy)
algorithm as well as its extended version (termed ExEDDy) are analyzed by means of numerical simulations. 
As a result it is found that while EDDy exhibits a rather poor perfomance, ExEDDy performes
considerably better and hence constitutes a significant improvement over EDDy.
Consequently, the results are contrasted with those
obtained for a block cipher that implements the encryption/decryption dynamics by means of
reversible cellular automata.
\end{abstract}


\end{frontmatter}

\section{Introduction}
\label{sect:Intro}

{\emph{Cryptography}} (Greek: {\emph{krypt\'os}} = hidden/secret, {\emph{gr\'aphein}} = writing) 
is the science of keeping information secure, e.g.\ for the purpose of secret communication \cite{HAC2010}.
In recent years it was realized that elementary, computer-science or physics motivated systems that exhibit a 
complex behavior allow for the design of cryptosystems, 
i.e.\ encryption algorithms (also termed \emph{ciphers}) and their decryption counterparts.
Among those ``applied science'' encryption algorithms one can distinguish \emph{stream ciphers} that encrypt a given
message one bit (or symbol) at a time \cite{wolframCryptography1985,baptista1998}, and \emph{block ciphers} that operate on blocks of, say, $N$ bits
\cite{ohira1998,ohira1999,seredynski2004}. 

The basic task in cryptography is two-fold: (i) the {\emph{encryption}} process, wherein a 
\emph{plaintext} message is scrambled using a certain cipher and \emph{key}. The key
specifies the precise way a given plaintext is transformed by means of the cipher. 
The result of the encryption procedure is a \emph{ciphertext} message.
(ii) the \emph{decryption} process, where the ciphertext is transformed into the 
initial plaintext message by use of the corresponding decryption algorithm and
a proper key. Note that, depending on the precise cryptography-mode, there is a 
subtle difference regarding the key: 
in \emph{symmetric-key cryptography}, where one deals with a single-key cryptosystem,
the same key is used during encryption/decryption. In \emph{public-key cryptography}, 
signifying a two-key cryptosystem, the encryption key is public and the decryption key is 
kept secret. 
A statement on the security of a cryptographic method is provided by ``Kerckhoffs principle''.
It states that a cryptosystem should be secure if everything about the cryptographic method
is known, except the key. 
Here, secure means that it is difficult or even impossible to obtain 
the plaintext from the ciphertext.
In this regard, a characteristic desirable for any cipher is that a change in plaintext or key,
as tiny as it may be, will result in a notable change in the ciphertext. More precise, changing 
one randomly chosen bit in either, plaintext or key, should invert nearly half of the bits of
the ciphertext. This \emph{avalanche property} was suggested by Feistel in 1973 \cite{feistel1973}.

Subsequently, only symmetric-key block ciphers designed
to process binary plaintext messages will be considered. As regards this, the bulk of the presented article
is dedicated to a scrambling scheme that relies on 
the paradigm of ``encryption via delayed dynamics'' (termed EDDy), see \cite{ohira1999}, as 
well as an extended variant thereof (signified as ExEDDy), see \cite{ohira1998}. 
In Ref.\ \cite{ohira1999} it was shown, how an exemplary initial state evolves with 
time and found that encoded states at different iteration times of the encryption 
dynamics are rather uncorrelated.
Further, the effect of a slight modification of the key 
on the ciphertext was illustrated for one plaintext message and concluded that 
a ``nearly'' correct guess of the
key does not lead to a sufficient overlap between decoded ciphertext and plaintext upon 
decryption so as to guess the initial message.
As a weakness of the method it was mentioned that a change 
of multiple bits in the plaintext propagates to the ciphertext as
a simple superposition of changes due to each changed bit.
In Ref.\ \cite{ohira1998} the encryption dynamics was modified to some extend, and
it was observed that a minor change in key leads to rather uncorrelated ciphertext messages 
and that a change of multiple bits in plaintext does  not propagate to the ciphertext as
a simple superposition of changes due to each changed bit.

\begin{figure}[t!]
\centerline{
\includegraphics[width=1.0\linewidth]{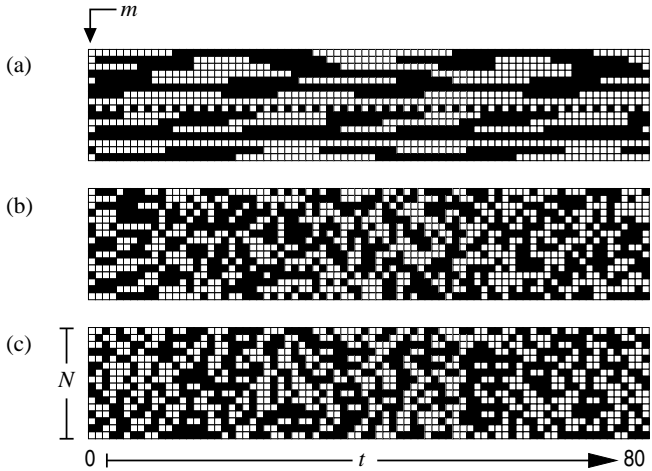}} 
\caption{
Illustration of the encryption dynamics for the cryptosystems under consideration.
All three ciphers were used to encode the same $N=16$-bit plaintext message $m$ (leftmost column) for
a number of $80$ iteration steps (bits with value $1$ are shown as nonfilled squares).
(a) Encryption via delayed dynamics (EDDy) cipher, (b) extended EDDy (i.e.\ ExEDDy) cipher
using the same key as in (a), and (c) reversible cellular automaton (ReCA) cipher using 
a randomly generated $32$-bit update rule.
\label{fig1}}
\end{figure}

The presented article addresses the question of performance of the two ciphers using numerical 
simulations. Basically speaking, the encryption algorithms
are applied to binary plaintexts of $N=4\ldots2048$ bits in length and it is checked
whether they exhibit the avalanche property.
Albeit there are more strict criteria to quantify the security of ciphers, I will solely consider 
the avalanche property. As it appears, EDDy does not exhibit the avalanche property.
Even worse, it also lacks \emph{completeness}. Completeness requires that for every possible key, every
ciphertext bit must depend on all plaintext bits, not only a proper subset of those. 
Further, I present clear evidence that ExEDDy outperforms EDDy.
So as to put the arguments along the plot of the presented article on solid ground,
a third, only recently introduced, cryptosystem based on reversible 
cellular automata (referred to as ReCA) as presented in \cite{seredynski2004} is considered and
the results obtained therein are partly reproduced to contrast those for (Ex)EDDy. 
The encryption algorithms considered here all work in iterative manner. As regards this,
Fig.\ \ref{fig1} gives a qualitative account of the difference in the respective
encryption dynamics.

Note that only few is known about the performance of reversible cellular automata based block 
ciphers \cite{seredynski2004} and even less is known on the performance of delayed dynamics
based ciphers such as EDDy and ExEDDy. To the best knowledge of the author, the presented
article comprises a first attempt to quantify how well the delayed dynamics based ciphers
EDDy and ExEDDy perform upon encrypting plaintexts of $N=4\ldots 2018$ bits in length.

The remainder of the presented article is organized as follows.
In section \ref{sect:algorithms}, the considered cryptosystems 
are introduced and illustrated in more detail.
Section \ref{sect:results} contains the numerical results obtained
via simulations for plaintext of different size and section \ref{sect:conclusions} concludes with 
a summary. 
A more elaborate summary of the presented article is available at the {\emph{papercore database}}
\cite{papercore}.

\section{Encryption algorithms}
\label{sect:algorithms}

Dynamical rules that describe the evolution of simple system often
induce complex behavior with characteristics that are desirable 
for an encryption scheme \cite{hayes1994,baptista1998,alvarez1999}.
In this regard, the remainder of this section describes three 
cryptographic methods that rely on iterative dynamic procedures, 
exhibiting complex behavior, used to scramble a binary input message
so as to hide its information content.

\subsection{EDDy: encryption via delayed dynamics}
\label{subsect:algorithms_EDDy}

Based on the paradigm of delayed dynamics, a block cipher for
encryption of binary data, here termed EDDy (as in ``encryption with delayed dynamics''), was proposed \cite{ohira1999}.
EDDy is a symmetric-key cryptosystem that works in an iterative manner and as such it requires
the following input:
(i)  a binary plaintext message $\myvec{m}=\myvec{s}(0)$, consisting of a sequence of $N$ bits $s_i(0)$ (taking values $\pm 1$), and
(ii) a key $\myvec{K}=(\myvec{P},\myvec{\tau},T)$, used during the encryption/decryption process.
The component $\myvec{P}$ signifies a permutation obtained from the sequence $(0,1,\ldots,N-1)$, 
$\myvec{\tau}$ represents a delay-time vector of length $N$, where $\tau_i\in[1,\tau_{max}]$ is an integer
delay time associated with bit $i$, and $T$ denotes the number of iterations carried 
out by the encryption algorithm. 
Note that the key required by EDDy is rather large: if we aim to encrypt a $N$-bit plaintext 
message and agree to use an integral data type that requires $M$-bits in 
order to represent the components of $P$ and $\tau$ as well as $T$, then the key has a length of $M(2N +1)$ bits
(bear in mind that the common $C/C_{++}$- data type {\tt int} has $32$ bits).
Now, from the point of view of an adversary that 
plans a cryptographic attack on EDDy, assume $\tau_{\rm max}$ is known. If the attack merely consists
in guessing a key, then there are as much as $(N!)(\tau_{\rm max})^N$ possible choices for a pair of $P$ and $\tau$, 
as well as a guess for $T$, left. Hence, the key space of EDDy is enormous.
For comparison: the \emph{data encryption standard} (DES) which is restricted to a blocksize of $64$ bits employs a $56$ bit key.
For fixed $N=64$ and $\tau_{\rm max}=10$, there are as much as $(N!)(\tau_{\rm max})^N \approx 2^{500}$ keys for the EDDy scheme.

Based on the ingredients above, the encryption dynamics read: 
\begin{align}
s_i(t)=-s_{P_i}(\max(0,t-\tau_i)) \label{eq:EDDy_cipher}
\end{align}
Therein, $P_i$ and $\tau_i$ are the $i$th elements of $\myvec{P}$ and $\myvec{\tau}$, respectively.
One iteration of the cipher consists in updating each bit once.
The ciphertext $\myvec{c}=\myvec{s}(T)$ is obtained when 
the exact number of iterations $T$, predefined in the key, are carried out. 
Decryption is performed by using the same key $\myvec{K}$ and applying the reverse dynamics 
$s_k(t)=-s_{u}(t+\tau_u)$, where $P_u=k$. So as to facilitate a proper decryption, 
the sender needs to communicate a sequence of at least $\tau_{\rm max}+1$ successive encoded states,
i.e.\ the ``full'' ciphertext $\myvec{c}_{\rm F}$, to the receiver.
An exemplary encryption procedure for a plaintext of $8$-bit in length is shown in Fig.\ \ref{fig:encryptionDynamics}(a).
In effect, the dynamics defined by the updating rule Eq.\ \ref{eq:EDDy_cipher} induces a linear dynamics. The state of a particular bit, say
bit $i$, is related to the state of exactly one other bit with integer identifier $j=P_i$ at some preceding time. 

\subsection{ExEDDy: extended variant of encryption via delayed dynamics}
\begin{figure}[t!]
\centerline{
\includegraphics[width=1.\linewidth]{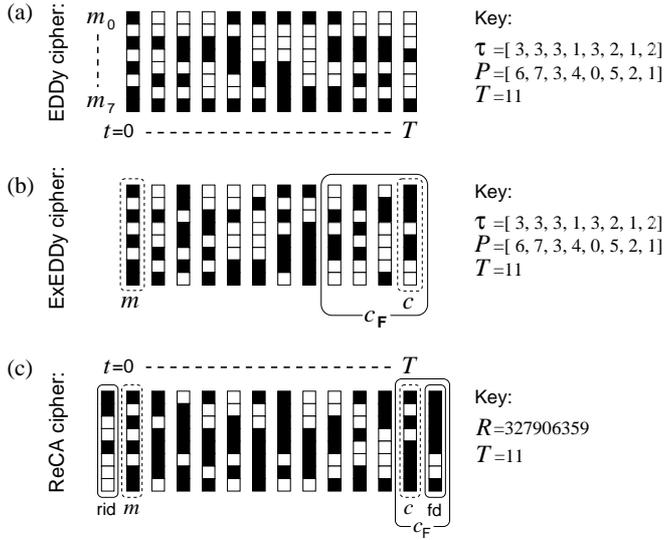}} 
\caption{
Illustration of the encryption process for a $8$-bit plaintext message 
$m=[-1,1,-1,1,-1,1,-1,-1]$ (bits in state $1$ are given by nonfilled squares) 
and key $K=(P,\tau,T)$ in case of (a) the EDDy cipher, and (b) the ExEDDy cipher.
In the latter figure, plaintext $m$, ciphertext $c$ and full ciphertext $c_{\rm F}$ are indicated. 
Figure (c) shows the encryption process for the ReCA cipher using plaintext 
$m=[0,1,0,1,0,1,0,0]$ and key $K=(R,T)$.
\label{fig:encryptionDynamics}}
\end{figure}

The extended variant of the EDDy cipher uses the same input information as above, i.e.\ 
a binary plaintext message $m=s(0)$ and a key $K=(P,\tau,T)$. Using the same notation
as above, the ExEDDy cipher reads
\begin{align}
s_i(t+1)= s_{P_i}(\max(0,t-\tau_i)) \cdot \theta\Big(\sum\nolimits_{i=0}^{N-1}s_i(t)\Big). 
\end{align}
Therein, $\theta(\cdot)$ is a step function satisfying 
\begin{align}
\theta(x)=\begin{cases}
  +1, & \text{if } x>0 \text{,}\\
  -1, & \text{if } x\leq0 \text{.}
\end{cases}
\end{align}
As opposed to the EDDy scheme, this causes a nonlinear dynamics since an arbitrary bit at time 
step $t$ (explicitly) depends on all $N$ bits at time step $t-1$.
Fig.\ \ref{fig:encryptionDynamics}(b) illustrates an exemplary encryption
procedure for the ExEDDy cipher.

\subsection{ReCA: encryption via reversible cellular automata}

Cellular automata (CA) are abstract computing models that work on a discrete 
space-time background. They exhibit complex dynamics and might be interpreted
as discrete structures that approximate differential equations \cite{vichniac1984}.
As such, CA highlight the close connection between the realms of computer science and
physics \cite{takesu1987,wolfram1983,wolframRandomness1985}. 
Further, there is a possibility to use CA as cryptosystems for the purpose of data 
encryption \cite{wolframCryptography1985,seredynski2004}. 

The most basic CA \cite{wolfram1983} consist of a circular array $s$ of, say, $N$ cells 
that can take values $s_i\in(0,1)$ for all $i\in(0,\ldots,N-1)$. 
The evolution of CA proceeds in discrete time steps $t$, where the values of the 
individual components $s_i(t)$ are updated synchronously according to a certain local update rule.
So as to evolve state $s_i(t) \to s_i(t+1)$, the local update rules $f$ considered in 
the presented article take into account the state of cell $i$ as well 
as the states of those cells that are located within a neighborhood of radius $r=2$ enclosing $i$ at time $t$, hence 
$s_i(t+1)=f[s_{i-2}(t),\ldots,s_{i+2}(t)]$ (note that the array of cells is circular, 
i.e.\ cells $0$ and $N-1$ are adjacent).
\begin{figure}[t!]
\begin{center}
\includegraphics[width=1.\linewidth]{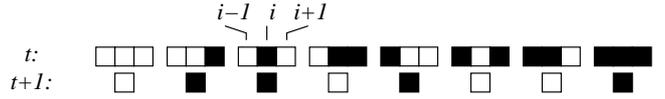}
\end{center}
\caption{
List of the $8$ possible configurations $s_{i-1}(t),s_i(t),s_{i+1}(t)$ 
for a radius $1$ CA, that might serve as input to the local update 
rule $f$.  The configurations are interpreted as $3$ digit binary
numbers (bits with value $1$ are shown as nonfilled squares) and 
sorted in decreasing fashion from left to right.
The $8$ solitary cells below cell $i$ indicate the state 
$s_i(t+1)=f[s_{i-1}(t), s_i(t), s_{i+1}(t+1)]$.
The respective $8$ bit sequence fully specifies the update rule.
In Wolfram's notation, the integer representation of the 
latter sequence is used as a name $R$ for the rule (here: $R=150$, 
which represents the most simple finite difference variant of 
the heat equation in modulo $2$ arithmetic \cite{vichniac1984}). 
\label{fig:8bitRule}}
\end{figure}  
Considering a neighborhood of radius $r$, there are an over all number of $2^{2r+1}$ 
different input configurations to $f$. Hence, the local rule for a one-dimensional radius $2$ 
CA is specified by a $2^{5}=32$-digit 
binary number. In turn, there are $2^{32}$ different CA rules for the specified setup.
A more simple example of a local rule for a $r=1$ CA is shown in Fig.\ \ref{fig:8bitRule}.
Therein, also the name convention for CA rules is detailed.
Note that in general, CA are not time reversal symmetric, a feature one would request
in order to set up a convenient dynamics for encryption as well as its reverse dynamics 
for decryption \cite{seredynski2004}.
Nevertheless, such basic CA might already be used as 
stream ciphers \cite{wolframCryptography1985,wolframRandomness1985}.
It is important to note that there is a possibility to transform any (non-reversible) rule into
a time reversal invariant rule. For a CA with a binary state alphabet and $r=2$ neighborhood, 
the recipe to achieve this simply reads:
$s_i(t+1) = ( f[s_{i-2}(t),\ldots,s_{i+2}(t)] -s_i(t-1)){\rm mod}(2)$ \cite{vichniac1984}. Note that the 
modified rule is second order in time, i.e.\ so as to start the dynamics of the CA one has to 
provide a configuration $s(t=0)$ and its predecessor $s(-1)$. The intriguing feature
of time reversal invariance is: a sequence of configurations can be obtained in reverse 
order by means of the same rule $f[\cdot]$, by reversing the last two configurations.
In effect, this is the core idea of encryption via reversible CA (ReCA) \cite{seredynski2004}.

In order to use ReCA for the purpose of data encryption, one has to supply two ingredients:
(i) the input data, where $s(0)=m$, and some random initial data (rid) taken
as $s(-1)$ so as to be able to start the ReCA dynamics.
(ii) a key that governs the encryption dynamics, given by an updating rule that specifies 
the evolution of the state $s(t)\to s(t+1)$ and a number $T$ of iteration steps, after which 
the ciphertext is obtained.
Note that for such a ReCA cipher, the full ciphertext consists of two sequences: $s(T)$ which 
is identified with the ciphertext $c$, and the final encrypted data (fd) $s(T+1)$.
An exemplary encryption process using a ReCA algorithm is illustrated in Fig.\ \ref{fig:encryptionDynamics}(c).
Decryption using ReCA is particularly simple:
upon setting $s^\prime(0)=s(T)$, $s^\prime(-1)=s(T+1)$,
and using the same key as during encryption, it will hold that $s^\prime(T)=m$.
The configuration $s^\prime(T+1)$ contains the random initial data and as such, 
it is not of interest after decryption. However, it might be used as an additional 
degree of freedom: upon using a particular key, a given plaintext message can be 
scrambled into numerous ciphertexts, depending on the random initial data.
So as to realize secure communication using this approach, there are some subtleties 
concerning the communication of the finial encrypted data, see Ref.\ \cite{seredynski2004}. 
However, the focus here is on the principal dynamics of the encryption schemes.

The subsequent section presents the results of the numerical simulations and 
attempts to give some more intuition on the principle dynamics on which the 
considered encryption algorithms are based.
%

\section{Results}
\label{sect:results}

In the presented section, the encryption process, wherein a plaintext message $m$ is 
scrambled using a certain cipher $F$ (either EDDy, ExEDDy or ReCA) and key $K$, is put 
under scrutiny. The result of the encryption procedure is a ciphertext message $c$, 
and the process might be written as $c=F_K(m)$.
In order to probe the avalanche criterion (the precise definition of the avalanche 
criterion is given below in Subsect.\ \ref{subsect:results_avalanchePar}) for a given 
cipher $F$, a three step procedure is adequate:
Obtain a random key $K$ (valid for $F$), then 
  (i) generate a random binary plaintext message, $N$ bits in length, and obtain 
      the corresponding ciphertext $c$,
 (ii) flip one bit, say bit $m_i$, of the plaintext message to obtain a perturbed 
      plaintext $m^{(i)}$ and obtain the corresponding ciphertext $c^{(i)}$,
(iii) compute the {\emph{Hammingdistance}} $d_{\rm H}(c,c^{(i)})$ between the two ciphertexts, 
      defined as the number of differences between the components of $c$ and $c^{(i)}$. 
      For binary sequences $c$ and $c^{(i)}$ of $N$ bit length and using Boolean notation, 
      the Hammingdistance may be written as 
\begin{eqnarray}
d_H(c,c^{(i)})=\#_1 (c \oplus c^{(i)}).
\end{eqnarray}
Therein $\oplus$ denotes the logical xor-operation, and $\#_1(c)$ signifies the number of 
occurrences of the digit $1$ in the binary representation of $c$.

For a given plaintext of length $N$, the results are then averaged over different plaintexts $m$ and 
keys $K$. In this regard, for $N\leq 10$ and for a particular key, it is feasible to consider 
the full input alphabet $\mathbb{Z}_2^N$ consisting of $2^N$ distinct plaintext messages. 
Here, a number of $200$ different keys was taken into account in order to compute averages
$\langle d_H(c,c^{(i)})\rangle_{m,K}$. 
For $N>10$, tuple consisting of a plaintext and a key were sampled uniformly at random
from among all possible choices. Further, the ``equiprobable ensemble'', featuring binary messages 
with probability $p=1/2$ for bits in state $1$, was considered. In this case, a number of $400$ 
plaintext/key pairs was used so as to compute averages.

In the remainder of the presented section, the dynamics of the encryption prosess is considered. 
In this respect, in the first subsection evidence is collected on whether plaintext and
ciphertext blocks, obtainded via the different encryption dynamics, are statistically independent.
For this purpose, a $n$-block frequency test 
(note that the frequency test carried out above resembles the monobit frequency
test suggested by NIST \cite{NIST} to check the randomness of binary sequences.
Here, instead of monobits, bytes are considered.) is employed, a compression test is performed, 
and a measure of (structural) complexity for sequences of encoded states is considered.
The respective subsection is closed by a simple graphical correlation analysis regarding
successive encoded states.
The subsequent subsection explains and quantifies the avalanche parameter for the considered
ciphers. Therefore, the evolution of the plaintext is considered and the stationary behavior of the 
avalanche parameter is examined. The presented section concludes with a note on completeness 
for the different encryption schemes.

\subsection{Dynamics of the encryption processes for the considered ciphers}
\label{subsect:results_dynamics}

A given cryptosystem provides \emph{perfect secrecy}, if the plaintext and ciphertext  
blocks processed by the cipher are statistically independent \cite{HAC2010}.
Here, the principle dynamics of the encryption process is considered by 
monitoring the evolution of plaintext messages during the encryption process. 
Therein, the aim of the presented subsection is to employ various tests on the 
three ciphers introduced above, to check if the sequence of encoded configurations 
exhibits obvious statistical regularities. To perform the necessary tests, 
a large number of up to $T\approx 10^4$ iteration steps are used with the respective
ciphers.
Albeit this might be infeasible for any practical cryptographic application, 
it is nevertheless useful as far as the principle dynamics of the algorithms 
is considered.
Further, so as to contrast the results on the three ciphers above, 
a $1d$ CA with $r=1$ neighborhood that implements Wolfram rule $30$ (denoted as $R30$) is considered.
$R30$ generates effectively random bit sequences and might be used as a stream cipher for 
cryptographic purposes, see Refs.\ \cite{wolframCryptography1985,wolframRandomness1985}. As such, it has already 
passed a series of statistical tests \cite{wolfram1986} and is solely considered for the purpose of comparison.

\begin{figure}[t!]
\begin{center}
\includegraphics[width=1.\linewidth]{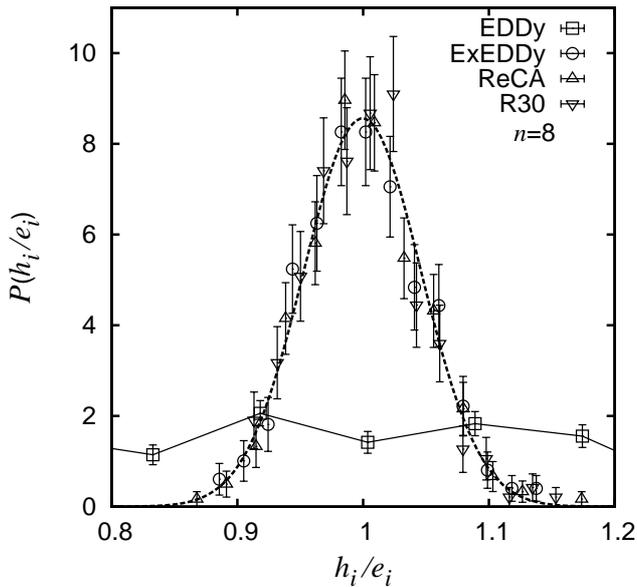}
\end{center}
\caption{
Result of the $n$-block frequency analysis. The pdf
shows the distribution of the ration of frequencies 
$\{h_i/e_i\}_{i=0}^{2^n-1}$ for $n=8$. 
The distributions resulting from ExEDDy, ReCA and $R30$ 
dynamics compare well to a Gaussian distribution with 
mean $\mu=0.999(2)$ and width $\sigma=0.047(1)$ (obtained
from a fit to the ReCA data and illustrated by 
the dashed line).
This already leads to expect that the histogram of 
observed frequencies is rather ``flat''.
The distribution related to the dynamics of the EDDy 
cipher suggests that for EDDy, the above conclusion 
is certainly not appropriate.
\label{fig:pdf_frequencies}}
\end{figure}  
\paragraph{$n$-block frequency test for configurations during encryption}
At first, the statistics for a sequence $s(t)$ of encrypted binary configurations,
obtained during the encryption process, is put under scrutiny. The basic 
question is whether the resulting statistics is compatible with that for 
sequences of effectively random configurations. 
A proper tool to check this is a $n$-block frequency test. 
This test can be cast into the following three step procedure:
(i)   slice the $N$ bit input configuration $s(t)$ for a given step $t$ 
      into a number of $M=\lfloor N/n \rfloor$ $n$-bit blocks
      $\{b_i\}_{i=1}^M$, and discard all remaining bits. 
      Then, follow the ``evolution'' of the input configuration up to $t_0+\Delta t$
      iteration steps, where the configurations obtained during the first $t_0=N$ iteration steps 
      are skipped (the value of $t_0$ is picked rather arbitrarily, it is ment to 
      eliminate possible effects of the initial ``short-time'' dynamics).
      Represent each $n$-bit block $b_i$ 
      by means of its integer decimal value ${\rm Id}[b_i]\in (0,\ldots,2^n-1)$,
      and accumulate the resulting integers in a histogram
      $\{h_i\}$ having $2^n$ bins with id $i\in (0,\ldots,2^n-1)$. 
      The bins then 
      contain the \emph{observed frequencies} $h_i$ associated to the integer numbers $i$. 
      For effectively random sequences one would expect to find each integer id with probability
      $2^{-n}$, hence the \emph{expected frequencies} are 
      simply given by $e_i = N_{\rm samp} \times 2^{-n}$ wherein $N_{\rm samp}=M \Delta t$.
      Now, if the encrypted configurations exhibit
      the statistics of random binary sequences, the histogram $\{h_i\}$ should
      be ``flat'' with each bin having the same expected frequency.
      Consequently the null hypothesis reads: ``The sample follows a 
     uniform distribution''.
(ii)  Probe the $\chi^2$ statistics, and, so as to check the null-hypothesis 
      compute the associated $p$-value, see Ref.\ \cite{practicalGuide2009}.
(iii) Finally, reject the hypothesis if $p\leq 0.01$. I.e., if $p>0.01$, the 
      sample follows the assumption that the histogram is flat.
The $\chi^2$-test described in steps (ii) and (iii) above was carried out for $N=128$-bit input sequences 
regarding four different dynamics given by the EDDy, ExEDDy, and ReCA cipher, as well as 
the $R30$ dynamics.
In the analysis, $\Delta t=10^3$ and blocks of $n=8$ bits in length, 
i.e.\ \emph{bytes}, where considered. 
The results of the analysis are listed in Tab.\ \ref{tab:tab1} and probability density 
functions (pdfs) for the ratio of the observed/expected frequencies related to the four different dynamics are
shown in Fig.\ \ref{fig:pdf_frequencies}. To summarize the findings: while the configurations
obtained using the ExEDDy, ReCA, and $R30$ dynamics exhibit statistics compatible
with those of random sequences, the computed $p$-values clearly suggest that EDDy gives rise to 
seemingly nonrandom sequences of configurations during encryption.
%
\begin{table}[b!]
\caption{\label{tab:tab1}
From left to right: 
Type of dynamics,
$\chi^2/{\rm dof}$ and $p$-values,
resulting from the $n$-block frequency analysis, 
compression factor $\kappa$ obtained using a simple data compression test,
and minimal number $N_{\rm ps}$ of pair substitution steps needed to 
transform a sequence of bit states into a constant sequence. For the
latter quantity, the values in the braces list the standard deviation.
} 
\begin{tabular}[c]{lllll}
\hline\noalign{\smallskip}
 	& $(\chi^2/{\rm dof},p)_{n=8}$  & $\langle \kappa \rangle$ & $\langle N_{\rm ps} \rangle$ \\
\noalign{\smallskip}\hline\noalign{\smallskip}
EDDy   	& (6.24, $10^{-197}$) 	& 0.1378(4) & 74(35) \\
ExEDDy 	& (0.87, 0.93) 	& 0.1714(7) & 218(34) \\
ReCA   	& (0.99, 0.53) 	& 0.17365(9)& 217(13) \\
$R30$ 	& (0.87, 0.94) 	& 0.17378(2)& 225(4)  \\ 
\noalign{\smallskip}\hline
\end{tabular}
\end{table}

\paragraph{An oversimplified compression test for the sequences of encoded states}
Besides checking whether the sequence $s(t)$ of encrypted configurations
exhibits the same statistics as truly random sequences,
it is also of interest to look for structural regularities. 
As regards this, here, an oversimplified compression test 
using the convenient Unix-tool {\tt gzip} is performed. 
As a prerequisite for the test, for each of the three encryption algorithms and 
the $R30$ dynamics, a number of $100$ plaintext messages
($128$ bits in length) is evolved for $10^3$ iteration steps. Again, $R30$ is considered for comparison only.
The evolution of each plaintext is stored in a single file. Prior to 
compression, the individual files have a size of, say, $B_0$ bits.
The {\tt gzip} data-compression tool uses the Lempel-Ziv (LZ77) algorithm \cite{lempel1977},
so as to reduce the size of the supplied data. Loosely speaking, 
LZ77 works by finding sequences of data that are repeated and 
exploits those ``patterns'' to perform the compression process.
After compression via {\tt gzip}, the size of a given file is reduced to $B_{\tt gzip}$.
Then, the compression factor $\kappa=B_{\tt gzip}/B_0$ for each file is computed
and its mean and standard error are listed in Tab.\ \ref{tab:tab1}.
As it appears, $\langle \kappa \rangle$ is approximately equal for the ExEDDy, ReCA and
$R30$ data, while it assumes a slightly smaller value for the EDDy data.
This indicates that, compared to the other encryption schemes,
EDDy exhibits a larger degree of structural regularity.

\begin{figure}[t!]
\begin{center}
\includegraphics[width=1.\linewidth]{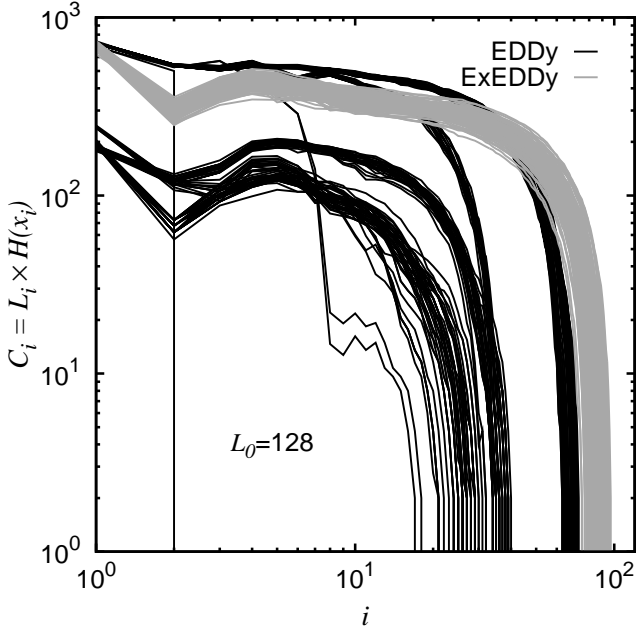}
\end{center}
\caption{
Results for a repeated application of the pair substitution process on 
bit strings, obtained by monitoring the states of
the individual bits of a 128 bit plaintext message during an exemplary encryption process
that took $10^3$ encryption steps.
Hence, there are 128 individual bit strings (corresponding to the 128 lines in 
the figure), each of which is initially 
$10^3$ symbols in length.
The figure shows the parameter $C_i$ which might be used 
to indicate whether a constant sequence is obtained during
iteration (in that case, it holds that $C_i=0$).
As discussed in the text, the application of the NSRPS algorithm
allows to define a measure of complexity associated to a given 
sequence of symbols.
\label{fig:pairSubs}}
\end{figure}  
\paragraph{A measure of the degree of randomness for the sequences of encoded states}
So as to elaborate on the degree of structural regularity, 
data compression algorithms can also be used to estimate further 
statistical properties of a given sequence $x$ of symbols. 
E.g., one might design a measure for the \emph{algorithmic randomness} 
associated to $x$ via non-sequential recursive pair substitution
(NSRPS) \cite{jimenezMontano2002,benedetto2006,nagaraj2011}.
In order to illustrate the NSRPS method, consider a sequence $x_0$ 
of symbols, composed of characters 
stemming from a finite $q$-symbol alphabet $A=\{a_i\}_{i=0}^{q-1}$. 
An elementary step of the NSRPS algorithm, referred to as \emph{pair substitution},
might be illustrated as two step procedure:
(i) For a given sequence $x_0$, determine the frequency of all 
(ordered) pairs $e\in A\times A$ with respect to $x_0$ and identify $e_{\rm mfp}$, 
i.e.\ the most frequent (ordered) pair
of symbols. If the most frequent pair is not unique, signify one of them as $e_{\rm mfp}$. 
(ii) Construct a new sequence $x_1$ from $x_0$ wherein each 
full pattern $e_{\rm mfp}$ is replaced by a new symbol $a_q$ (for this purpose, $x_0$ is scanned
from left to right), and augment $A$ with $a_q$. The new sequence
is shorter in length and the size of its associated alphabet is incremented by 1.  
This pair substitution step might be executed iteratively, until the representation
of the sequence requires a single character, only. Such a sequence has 
zero information entropy 
\begin{align}
H(x)=- \sum_{i=0}^{q-1} \frac{n_i}{N} \log_2\Big(\frac{n_i}{N}\Big), \label{eq:infEntropy}
\end{align}
wherein $n_i$ specifies the observed frequency of symbol $a_i$ in the sequence $x$ of length $N$,
and is called a \emph{constant} sequence.
Here, as a measure of algorithmic randomness associated to the initial sequence $x_0$, the minimal 
number of pair substitution steps $N_{\rm ps}$, needed to 
transform $x_0$ to a constant sequence with $H(x_{N_{\rm ps}})=0$, is considered.
In order to facilitate intuition, consider the $6$ bit sequence $x=110010$. 
An iterative application of the pair substitution step on $x$ yields the result:
\begin{center}
\begin{tabular}[l]{cllll}
step $i$ & $x_i$	& $H(x_i)$ 	& $e_{\rm mfp}$ & $a_q$  \\
0	& $110010$	& $1.0$		& $10$		& $2$	\\	
1	& $1202$	& $1.5$		& $02$		& $3$	\\
2	& $123$		& $1.58$	& $23$		& $4$	\\
3	& $14$		& $1.0$ 	& $14$		& $5$	\\
4	& $5$		& $0.0$ 	&		&	\\
\end{tabular}
\end{center}
I.e., after $N_{\rm ps}=4$ steps, the NSRPS algorithm terminates 
for the constant sequence $x_4=5$.
For the slightly different sequence $y=101010$ (also with initial entropy $H(y_0)=1$),
the NSRPS algorithm terminates after a single pair substitution step, where 
the constant sequence reads $y_1=222$.
One may now conclude that sequence $x$ exhibits a higher degree of randomness than sequence $y$, since
the NSRPS algorithm requires more elementary pair substitution steps in order 
to arrive at a constant sequence.
This is in accord with intuition, since, in contrast to sequence $x$, $y$ shows a regular pattern.

In Fig.\ \ref{fig:pairSubs}, results for a repeated application of the pair substitution process on 
$128$ individual bit strings, each of which is $10^3$ bits in length, are shown. 
The bit strings are obtained as follows: a $128$ bit plaintext is prepared and subsequently encrypted 
using $10^3$ encryption steps. Each individual bit of the plaintext is monitored during the encryption
process so as to yield a bit sequence that is $10^3$ bits in length.
More precise, the figure illustrates the evolution of the parameter $C_i=L_i \times H(x_i)$ for each of 
the 128 individual bit strings (each of which initially is $10^3$ symbols in length), wherein $L_i$ signifies
the length of the sequence at iteration step $i$.
Results are shown for the EDDy and ExEDDy ciphers, only.
Upon execution of the NSRPS algorithm, the parameter value $C_i=0$ might be used as a stopping criterion to 
detect whether a constant sequence is obtained \cite{nagaraj2011}.
Here, the application of the NSRPS algorithm
to each bit string sheds some light on the ``internal'' dynamics
of the encryption schemes.
Consider, e.g., the EDDy cipher: as evident from Fig.\ \ref{fig:pairSubs}, there are some bit strings that 
transform to constant ones almost immediately, while there are others that last for quite a large number 
of iteration steps.
\begin{figure}[t!]
\begin{center}
\includegraphics[width=1.\linewidth]{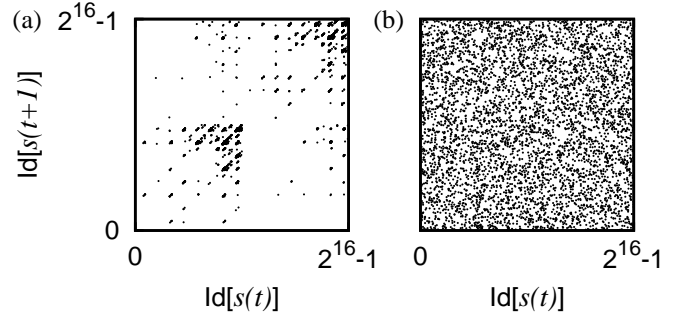}
\end{center}
\caption{
Typical result for a graphical correlation analysis for subsequent
configurations $s(t)$ and $s(t+1)$ obtained using the encryption dynamics
based on (a) EDDy, and (b) ExEDDy (using the same plaintext and key as (a)).
In the figure, a particular
binary configuration $s(t)$ is represented by means of its integer decimal value 
 ${\rm Id}[s(t)]$.
The results for the ReCA cipher look similar to those of ExEDDy and 
are not shown.
\label{fig:corrGraphical}}
\end{figure}  
The reason for this has an intuitive appeal:
during an encryption process using EDDy, the dynamics of 
a given bit is only coupled to those bits that are located 
along the same cycle in the cycle decomposition of the index-permutation 
$P$ that forms part of the key (see discussion below). 
Hence, the encryption dynamics of EDDy for a plaintext of a given size can be decomposed into 
independent dynamics taking place on the set of cycles in the cycle-decomposition of $P$.
Depending on the particular delay-times that characterize the bits in such a cycle, 
the respective dynamics might turn out to be rather regular.
This can most easily be seen for a fixed point $p_i=i$ of the permutation $P$: once the 
dynamics $s_i(t)=-s_i(\max(0,t-\tau_i))$ is kicked off, the state of the respective bit 
changes with period $\tau_i$.
For the simple case where $\tau_i=1$, the sequence of the state of that particular bit 
reads just $01010101\ldots$ (given that $s_i(0)=0$), and the NSRPS algorithm will 
terminate after a single step. Similarly, a delay time $\tau_i=2$ might yield the sequence
$00110011\ldots$, keeping the NSRPS algorithm busy for 3 steps.
Somewhat ``less regular'' patterns are found for cycles that are larger in length.
Consequently, a widespread distribution of the pair substitution steps $N_{\rm ps}$,
needed to obtain a constant sequence, is observed. The average number of such pair substitution
steps, taking into account 200 different 128 bit plaintexts, that where evolved for 
$10^3$ iteration steps each, is listed in Tab.\ \ref{tab:tab1}.
The results for the ExEDDy cipher are completely different: the fates of the individual 
bit sequences upon application of the NSRPS algorithm appear to be less diverse, 
see Fig.\ \ref{fig:pairSubs}. 
In comparison to the results obtained for EDDy, this is manifested by a smaller
standard deviation associated to $\langle N_{\rm ps} \rangle$, as listed in Tab.\ \ref{tab:tab1}.
The results for the ReCA cryptosystem and $R30$ dynamics are listed in Tab.\ \ref{tab:tab1}, also.
They look similar to those obtained for ExEDDy and are thus not shown.

\paragraph{Graphical analysis of correlations between successive encoded states}
One may further wonder whether the sequence $s(t)$ obtained during encryption exhibits 
correlations with respect to, e.g., subsequent time steps. 
In order to check this, it is most simple to label a configuration at
a given time step $t$ by means of its integer decimal representation ${\rm Id}[s(t)]$ and 
to plot ${\rm Id}[s(t+\Delta t)]$ as function of ${\rm Id}[s(t)]$.
Such a graphical correlation analysis for the evolution of one random $16$ bit plaintext
message is illustrated in Fig.\ \ref{fig:corrGraphical}(a) and (b), following the encryption 
dynamics of the EDDy and ExEDDy ciphers, respectively. Therein, the correlation between subsequent 
configurations, i.e.\ $\Delta t=1$, was considered. 
As evident from the figure,
EDDy exhibits a statistically rather regular behavior, which might be attributed to 
its linear dynamics, whereas ExEDDy and ReCA exhibit no obvious regularities for $\Delta t=1$ 
(the results for the ReCA dynamics look similar to those of ExEDDy and are not shown). Further, 
the regular pattern generated by EDDy suggests a rather short cycle length for its dynamics.

\subsection{Quantifying the avalanche parameter for the considered ciphers}
\label{subsect:results_avalanchePar}
The avalanche criterion would require to check whether $\langle d_H(c,c^{(i)})\rangle_{m,K}\approx1/2$
for each $i\in(0,\ldots,N-1)$. Here, the bits are
thought of as equivalent entities, thus, in the remainder of the presented section, only one
flip for each plaintext is attempted. This means, step (ii) above yields a modified plaintext $m^\prime$
(where the flipped bit is chosen uniformly at random), 
for which $c^\prime=F_K(m^\prime)$ and it is only checked whether 
$\langle d_H(c,c^\prime)\rangle_{m,K}$ agrees with the desired value $1/2$ within errorbars. If the latter requirement is satisfied, the 
respective cipher is said to exhibit the avalanche property. 
Subsequently, the Hammingdistance, averaged over different plaintext messages $m$ and
keys $K$, i.e.\ $\langle d_H(c,c^\prime)\rangle_{m,K}$, is referred to as \emph{avalanche parameter}.

\begin{figure}[t!]
\begin{center}
\includegraphics[width=1.\linewidth]{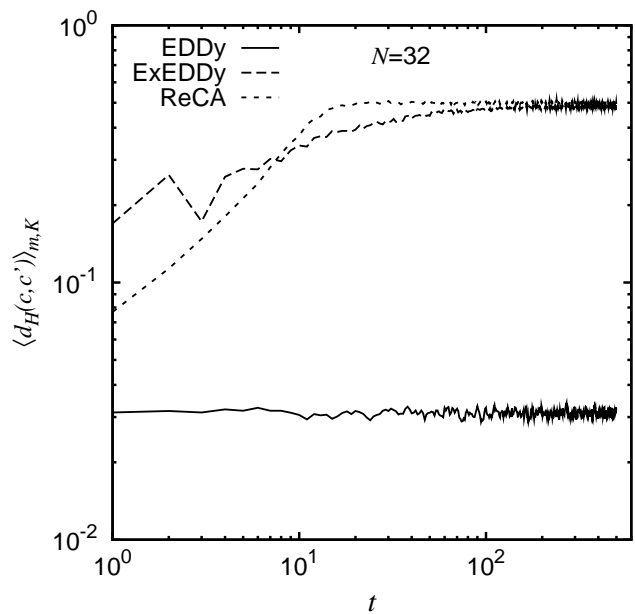}
\end{center}
\caption{
Results for the Hamming-distance $\langle d_{\rm H}(c,c^\prime) \rangle_{m,K}$
as function of the number of iteration steps $t$, obtained for the 
different encryption algorithms.
\label{fig:dHInitialStage}}
\end{figure}
\begin{figure}[t!]
\begin{center}
\includegraphics[width=1.\linewidth]{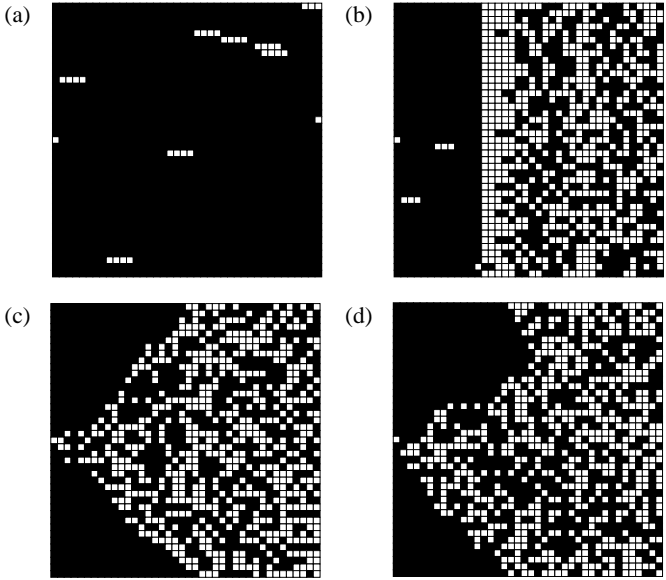}
\end{center}
\caption{
Exemplary plots that illustrate the evolution of the difference $c \oplus c^\prime$
between two ciphertexts $c=f_K(m)$ and $c^\prime=f_K(m^\prime)$,
where initially $m$ and $m^\prime$ differed only in the value of the centered bit.
Nonfilled squares signify bits for which the values in the two configurations are different.
Results for encryption dynamics based on (a) EDDy, (b) ExEDDy (using the same key as (a)), 
as well as
(c) isotropic spread, and (d) anisotropic spread of the difference obtained using the 
ReCA cipher. 
The length of the plaintext was $N=41$ and the evolution was followed over $40$ iteration steps of
the particular encryption dynamics.
\label{fig:hammingDist}}
\end{figure}

\begin{figure}[t!]
\begin{center}
\includegraphics[width=1.\linewidth]{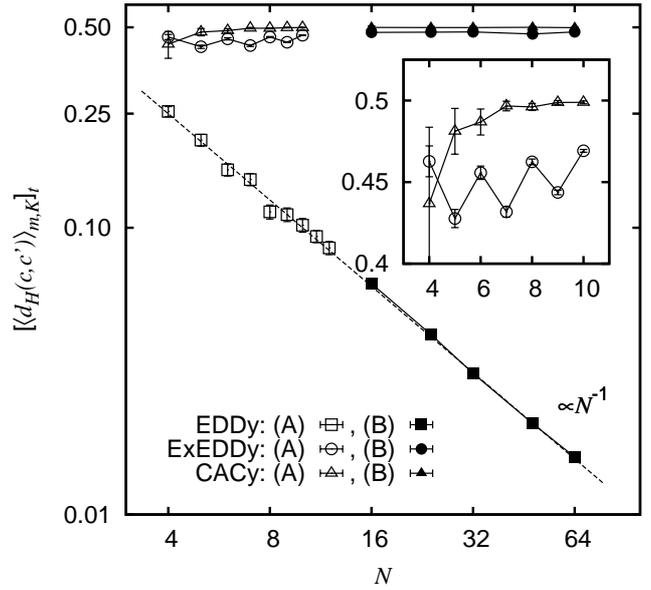}
\end{center}
\caption{
Avalanche parameter resulting from a flip of a single plaintext bit 
as a function of the plaintext bit-length $N$.
Nonfilled symbols 
indicate an average over the full input alphabet, while the filled
symbols refer to a stochastic sampling of plaintext messages. 
Error bars display the standard error associated 
to the data points.
The inset shows a zoom in on the ExEDDy and ReCA results for small 
plaintext lengths. 
\label{fig:avalanchePlaintext}}
\end{figure}

  
\paragraph{Evolution of the Hammingdistance upon modification of the plaintext}
Regarding the EDDy cipher, the Hammingdistance $\langle d_{\rm H}(c,c^\prime) \rangle_{m,K}$, interpreted
as a function of the iteration steps $t$ carried out by the encryption dynamics, fluctuates
around a value of $N^{-1}$ for all considered plaintext lengths $N$. Fig.\ \ref{fig:dHInitialStage} 
illustrates the results at $N=32$, where the distribution of the distance-values
$\langle d_{\rm H}(c,c^\prime) \rangle_{m,K}$, collected over an appropriate number 
of iteration steps,
follows a Gaussian distribution with mean value $\mu(d_{\rm H})\approx 0.031$ and 
width $\sigma(d_{\rm H})\approx0.001$ (not shown).
The argument for such a small mean value is quite plausible:
Consider, e.g., a $6$ bit plaintext, encrypted by using a key containing the index-permutation
$P=(3, 4, 2, 0, 5, 1)$. The respective cycle decomposition reads $(2)(3,0)(4,5,1)$. As 
a result, the ciphertext-bit $c_5$ is only affected by a finite number of plaintext-bits, namely bits 
$m_4$ and $m_1$. Now, if the value of a single bit in the plaintext is changed, say $m_5^\prime = -m_5$, the
difference in the evolution of the altered plaintext will be confined to the
cycle $(4,5,1)$ only. 
This is illustrated in Fig.\ \ref{fig:hammingDist}(a), where the difference
between two configurations $c$ and $c^\prime$, i.e.\ $c \oplus c^\prime$, 
during an encryption process using EDDy is monitored. 
Therein, the encryption processes was started at two plaintexts that differed only in the value of the center bit.
In the analysis, if the flipped bit is chosen uniformly at random, 
the length of the corresponding cycle is best fit by the expression 
$\langle \ell_{\rm c} \rangle = 0.49(1) (N-0.9(2))^{1.003(4)} \approx N/2$
(where the probability mass function for cycles with length $\ell$ behaves as
$n(\ell)\propto \ell^{-1}$, measured for permutations $P$ of length $N=128$; not shown).
Now, the length $\ell_{\rm c}$ of a particular cycle sets an upper bound for the 
Hammingdistance, giving rise to the restriction 
$\langle d_{\rm H}(c,c^\prime) \rangle_{m,K} \leq 1/2$.
Apparently, since the difference $c \oplus c^\prime$ typically stems from only a small number 
of bits contained in the respective cycle (see Fig.\ \ref{fig:hammingDist}(a)), it
appears that on average
$[\langle d_{\rm H}(c,c^\prime) \rangle_{m,K}]_t \approx N^{-1}$ for all $N$ considered 
(see Fig.\ \ref{fig:avalanchePlaintext} and discussion below).

As evident from Fig.\ \ref{fig:dHInitialStage}, the behavior of the ExEDDy cipher is 
completely different: the Hammingdistance initially increases but saturates at a value 
slightly below $1/2$.
Related to this, Fig.\ \ref{fig:hammingDist}(b) illustrates the difference between 
two configurations that initially differed only in the value of the center bit.
Considering a CA with a neighborhood of radius $r=2$, an initial change in 
plaintext, i.e.\ at time $t=0$, may affect at most $2rt+1$ cells at 
time $t$, implying that $d_H\leq (2rt+1)/N$. Therein, the value 
of $r$ defines an intrinsic ``velocity'', measuring the number
of cells per time-step, by means of which a change is transmitted
through the CA array. However, note that this is merely an upper bound, the 
actual velocity depends on the precise updating rule implemented by the CA 
and might also be anisotropic (cf.\ rules $30$ and $60$ for elementary
cellular automata \cite{weissteinElementaryCA}).
To illustrate this, Figs.\ \ref{fig:hammingDist}(c),(d) show the difference
between two configurations that initially differed only
in the value of the bit at the center, based on the ReCA cipher.
Then, for the ReCA dynamics one can expect an approximate linear 
increase of $d_{\rm H}$ during the initial stage of the encryption
dynamics. In deed, a fit using a functional 
form $\langle d_{\rm H}(c,c^\prime) \rangle_{m,K} = (u\cdot t+1)/N$ 
yields the result $u=1.112(1)$ at $N=256$. 
Analysis of the initial dynamics for further plaintext lengths $N=4,\ldots,512$ yield 
qualitatively similar results (not shown).
Hence, after $t \approx N/2$ 
time steps a change of one bit in plaintext leads to a distance
$\langle d_{\rm H}(c,c^\prime) \rangle_{m,K}\approx 1/2 $ between the ciphertext
corresponding to the perturbed/non-perturbed initial message, see Fig.\ \ref{fig:dHInitialStage}.
This finding compares well to the observation reported in Ref.\ \cite{seredynski2004},
where for a $r=2$ CA with $32$ ($64$) cells a minimum number of $19$ ($38$) iterations of the encryption
dynamics was recommended.

\paragraph{Stationary behavior of the avalanche parameter}
So as to quantify the stationary behavior of the avalanche parameter, it is useful
to distinguish the kind of {\emph{modification}} imposed on the encryption process.
In this regard, the presented analysis distinguishes between 
(i) a modification of the plaintext, where a single bit of the plaintext is changed (the bit is chosen 
uniformly at random), and 
(ii) a modification of the key. In the latter case, the focus is on the EDDy/ExEDDy ciphers only.

As regards (i), the avalanche parameter for the EDDy cipher (averaged over the iteration steps $t$) shows the clear scaling 
behavior $[\langle d_{\rm H}(c,c^\prime) \rangle_{m,K}]_t\propto N^{-1}$ for the full 
range of the considered plaintext bit-lengths $N$, see Fig.\ \ref{fig:avalanchePlaintext}.
This is in accord with the observation above, where the evolution of the Hammingdistance 
during an encryption process was considered.
The behavior of the ExEDDy cipher is strikingly different: the value of the avalanche
parameter, averaged over iteration steps $t\geq 200$, is only slightly smaller than $1/2$, 
see Fig.\ \ref{fig:avalanchePlaintext}. 
E.g., the numerical value $[\langle d_{\rm H}(c,c^\prime) \rangle_{m,K}]_t=0.482(5)$ attained at 
$N=32$ is only $3.5$ standard deviations below the desired value. 
Given the apparent similarity of the two encryption schemes, this is a considerable improvement over EDDy.
The ReCA based cipher appears to yield the best results: while there are slight deviations 
from the desired value for small plaintext sizes $N<9$ (e.g., at $N=8$ the numerical value
$[\langle d_{\rm H}(c,c^\prime) \rangle_{m,K}]_t=0.496(2)$ of the 
avalanche parameter is $2$ standard deviations below $1/2$)
the value of the avalanche parameter is compatible with $1/2$ for all values $N=9\ldots 64$
(e.g., at $N=64$ one finds $[\langle d_{\rm H}(c,c^\prime) \rangle_{m,K}]_t=0.498(3)$).
The results for the ReCA cipher where averaged over iteration steps $t\geq 350$.

\begin{figure}[t!]
\begin{center}
\includegraphics[width=1.\linewidth]{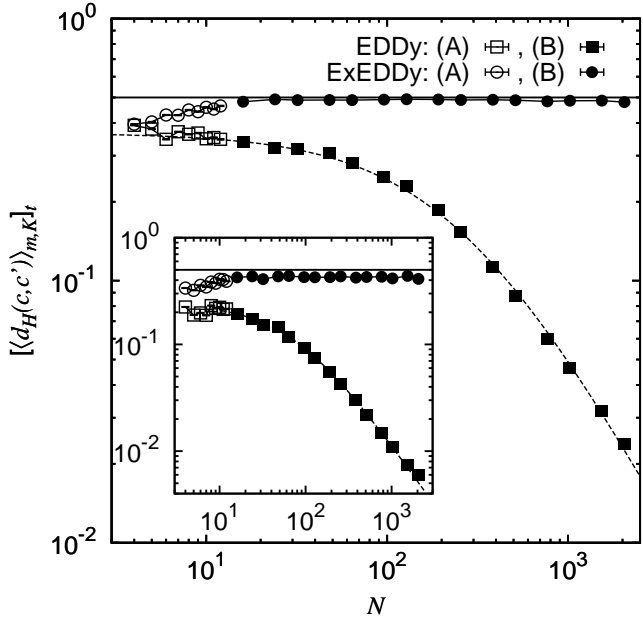}
\end{center}
\caption{
Avalanche parameter resulting from swapping two entries of the index-permutation $P$
 (illustrated by the main plot), and choosing a bit (uniformly at random) and 
replacing its associated delay-time by a new one (shown in the inset).
The abscissa indicates the length $N$ of the plaintext.
Nonfilled symbols indicate an average over the full input alphabet, while the filled
symbols refer to a stochastic sampling of plaintext messages. 
Error bars display the standard error associated 
to the data points (for the filled symbols, the symbol size is larger than the
error bars).
The solid horizontal line indicates a value of $[\langle d_{\rm H}(c,c^\prime) \rangle_{m,K}]_t=1/2$.
\label{fig:avalancheKey}}
\end{figure}  
Considering a modification of the key and focusing on the EDDy and ExEDDy cryptosystems, 
a further distinction is drawn between altering the permutation $P$ and the delay-time vector $\tau$.
Regarding the former modification, two randomly chosen entries are swapped, while in the 
latter case a single bit is chosen uniformly at random and the associated delay time is redrawn from 
the interval $[1,\tau_{\rm max}]$.
In both cases, the results for the EDDy cipher are best fit by powerlaw functions of the 
form $[\langle d_{\rm H}(c,c^\prime) \rangle_{m,K}]_t\propto (N+\Delta N)^{a}$, where 
$\Delta N\approx265$ and $a=-1.29(6)$ for a change in $P$, see Fig.\ \ref{fig:avalancheKey}, and 
$\Delta N\approx65$ and $a=-1.09(3)$ for a change in $\tau$, see inset of Fig.\ \ref{fig:avalancheKey}.
Consequently, in the limit of large $N$, the effect of a minimal change
in the plaintext on the ciphertext is negligible. 
The precise numerical values of
$[\langle d_{\rm H}(c,c^\prime) \rangle_{m,K}]_t$ in the range $N=24\ldots2048$ are about 
$2-5$ standard deviations below $1/2$.

For ExEDDy, however, results are completely different: 
upon swapping two random entries of the index-permutation, the avalanche parameter (averaged 
over iteration steps $t\geq 200$) is only slightly below $1/2$, see Figs.\ \ref{fig:avalancheKey}.
As regards this, the numerical values of
$[\langle d_{\rm H}(c,c^\prime) \rangle_{m,K}]_t$ in the range $N=24\ldots2048$ are about 
$2-5$ standard deviations below $1/2$.
Further, changing one randomly chosen delay-time yields an avalanche parameter (averaged over
the data for $t\geq 200$) clearly smaller than $1/2$, see inset of Fig.\ \ref{fig:avalancheKey}.
In conclusion, a modification of plaintext or index permutation leads to quite similar
numerical values for the associated avalanche parameters. In comparison, a modification of a 
randomly chosen delay time results in somewhat smaller values of the avalanche parameter.

\subsection{A note on completeness}
\label{subsect:results_completeness}
As pointed out above, the encryption dynamics of EDDy for a plaintext of a given size can be decomposed into 
independent dynamics taking place on the set of cycles in the cycle-decomposition of $P$.
As a result, the dynamics of a given bit is only coupled to those bits that are located 
in the same cycle of the cycle decomposition. This means that typically, a given ciphertext bit depends
on a few plaintext bits only.
Thus, the linear dynamics provided by EDDy lacks completeness (as explained in the introduction). 
The situation is different for ExEDDy: since each bit depends on the sum of states of all bits
in the previous time step, completeness is trivially satisfied.
Also, if the dynamics of the ReCA encryption algorithm is iterated for $t_{\rm min}\gtrapprox N/2$ 
time-steps, each bit in ciphertext might be affected by a given bit in the plaintext message.
Note that this argument was established from a statistical point of view (as discussed above), 
the precise value of $t_{\rm min}$ depends on the particular key used during the encryption process.
Further, bear in mind that, as discussed above, a lower bound on the respective time scale for 
a $r=2$ CA is given by $t=N/4$. Hence, ReCA exhibits completeness after a certain (key dependent) 
minimal number of time-steps.

\section{Conclusions}
\label{sect:conclusions}

In the presented article, the encryption process of three different cryptosystems for 
block-encryption, i.e.\ EDDy, ExEDDy (an extended variant of EDDy) and ReCA, is put under 
scrutiny. While EDDy and ExEDDy are based on the paradigm of delayed dynamics, ReCA relies 
on the updating rules provided by reversible cellular automata. Starting from an initial plaintext message, 
all three cryptographic schemes iteratively process the plaintext, yielding a sequence
of encoded configurations until, finally, the ciphertext is obtained.
Upon analysis of the statistical properties related to the sequence of encoded configurations,
it was found that the EDDy cipher gives rise to a statistically rather regular behavior: 
the sequence of encoded configurations does not exhibit the properties expected for
effectively random sequences. For the purpose of comparison, effectively
random binary sequences produced by a certain $1d$ CA, referred to as $R30$, were considered.
Further, EDDy does not satisfy the avalanche criterion (neither for a modification of plaintext
nor a modification of the key) and it lacks completeness.
In contrast to this, the ExEDDy cipher clearly outperforms EDDy in all test considered. 
The statistical properties of the sequence of encoded configurations are in good agreement 
with those of the ReCA cipher and the $R30$ dynamics. 
ExEDDy does not strictly satisfy the avalanche criterion but performs reasonably well.
Further, it features completeness. Given the apparent similarity of the algorithmic 
procedures that underlie both delayed-dynamics based ciphers, ExEDDy constitutes a striking 
improvement over EDDy.
Finally, the statistical properties of the sequences of encoded configurations that 
are obtained during an encryption process using the ReCA cipher are in excellent agreement with 
those obtained for the $R30$ dynamics. From the analysis of plaintexts with bit-size $N=4\ldots 64$ 
it is obvious that ReCA satisfies the avalanche criterion (in support of Ref.\ \cite{seredynski2004}, 
where $32$ and $64$-bit plaintexts are considered only). Further, the ReCA cipher shows
completeness after a key-dependent minimal number of iteration steps.

Another issue of interest from a point of view of application is the size of the full ciphertext, relative 
to the plaintext. As regards this, the three cryptosystems considered here lead to \emph{data expansion}.
As pointed out in subsect.\ \ref{subsect:algorithms_EDDy}, the key space of the delayed dynamics
based cryptosystems increases with the maximal delay time as $\propto (\tau_{\rm max})^N$, 
ensuring a high level of security for large $\tau_{\rm max}$. However, note the flipside 
to that coin: the minimal size of the full ciphertext that is necessary for decryption 
increases as $(\tau_{\rm max}+1)\times N$ and thus depends on the parameter $\tau_{\rm max}$ 
as part of the key.
In contrast to the delayed dynamics ciphers, the size of the full ciphertext for the 
ReCA cryptosystem is always $2\times N$.
To the knowledge of the author, the cryptographic schemes discussed in the presented 
article are of academic interest. The respective encryption dynamics are goverend by rather 
basic model dynamics that, without doubt, are interesting in their own right since 
they give rise to a very complex behavior and eventually exhibit properties that can 
be exploited so as to set up a working encryption algorithm. However, so as to implement a full
cryptographic method that allows for encryption and decryption of data, a rather 
large key is necessary and the ciphertext may turn out to be considerably larger in size than the 
plaintext (also known as \emph{data expansion}). 
These effects somehow limit the practicabillity of the discussed cryptographic schemes.


\section*{Acknowledgment}
OM would like to thank C. Norrenbrock and A. K. Hartmann for critically reading the manuscript.
Further, OM acknowledges financial support from the DFG (\emph{Deutsche Forschungsgemeinschaft}) under
grant HA3169/3-1. The simulations were performed at the GOLEM I cluster for scientific computing at 
the University of Oldenburg (Germany).

\bibliographystyle{model1-num-names}
\bibliography{literature_DDy.bib}
\end{document}